\documentclass{article}
\usepackage{epsf}

\newcommand{\eps}{\epsilon}
\newcommand{\intinf}{\ensuremath{\int_{-\infty}^{\infty}\!\!\! }}
\newcommand{\disorderop}{\ensuremath{
   \sum_{\alpha\neq\beta}^{n} \phi_{12}^{\alpha}\phi_{12}^{\beta} }}
\newcommand{\disop}[3]{\ensuremath{
   \sum_{#1\neq#2}^{n} \phi_{12}^{#1}(x_{#3}) \phi_{12}^{#2}(x_{#3}) }}
\newcommand{\disopb}[3]{\ensuremath{
   \sum_{#1\neq#2}^{n} \phi_{12}^{#1}(#3,y=0) \phi_{12}^{#2}(#3,y=0) }}
\newcommand{\disopbb}[3]{\ensuremath{
   \sum_{#1\neq#2}^{n} \phi_{12}^{#1}(#3,0) \phi_{12}^{#2}(#3,0) }}
\newcommand{\disopc}[2]{\ensuremath{
   \sum_{#1\neq#2}^{n} \phi_{12}^{#1}\phi_{12}^{#2} }}
\newcommand{\disopd}[2]{\ensuremath{
   \sum_{#1\neq#2}^{n} \varepsilon^{#1}\varepsilon^{#2} }}
\newcommand{\perturbation}{\ensuremath{
   \Delta_0 \intinf dx \disorderop}}
\newcommand{\sepr}[2]{\ensuremath{
   \mid x_{#1} - x_{#2} \mid < r}}
\newcommand{\sepri}[2]{\ensuremath{
   \int\limits_{\sepr{#1}{#2}}\!\!\!\!\!  }}
\newcommand{\twoptcorr}[2]{\ensuremath{
   <\phi_{12}(x_{#1})\phi_{12}(x_{#2})>_0}}
\newcommand{\twoptcorrno}[2]{\ensuremath{
   <\phi_{12}(x_{#1})\phi_{12}(x_{#2})>}}
\newcommand{\twoptcorrb}[2]{\ensuremath{
   \mid x_{#1} - x_{#2} \mid^{-1+\epsilon}}}
\newcommand{\abspow}[2]{\ensuremath{
   \mid #1 \mid^{#2}}}
\newcommand{\irrsum}[2]{\ensuremath{
   \sum_{\stackrel{{#1}_i\neq {#1}_j}{1\leq {#1}_i\leq (n-#2)}} }}
\newcommand{\irrsumb}[1]{\ensuremath{
   \sum_{\stackrel{{#1}_i\neq {#1}_j}{1\leq {#1}_i\leq n}} }}
\newcommand{\irrterm}[3]{\ensuremath{
   (\phi_{12}^{{#1}_1}-\phi_{12}^{n})\ldots
   (\phi_{12}^{{#1}_{#2}}-\phi_{12}^{n-(#2-1)})
   \phi_{12}^{{#1}_{M+1}}\ldots\phi_{12}^{{#1}_{#3}} }}
\newcommand{\OPEb}{\ensuremath{{\tilde{b}}_{NMn}}}
\newcommand{\OPEbb}{\ensuremath{{\tilde{b}}_{NM0}}}

\newcommand{\opprod}{\ensuremath{\prod_{\alpha=1}^{2N}
   \phi_{12}^{\alpha} (0,y)}}
\newcommand{\opproda}{\ensuremath{\prod_{\alpha=1}^{2N}
   \phi_{12}^{\alpha} (0,1)}}

\newcommand{\ph}[1]{\ensuremath{\phi_{12}^{#1}}}
\newcommand{\phtwo}[2]{\ensuremath{
   (\phi_{12}^{#1}-\phi_{12}^{#2})}}
\newcommand{\Q}[1]{\ensuremath{Q_{NMn}^{(#1)}}}
\newcommand{\I}[1]{\ensuremath{\mathcal{I}_{#1}}}

\newcommand{\Oirr}{\ensuremath{
   \mathcal{O}_{NMn}}}
\newcommand{\biOirr}{\ensuremath{
   \left(\intinf dx \Oirr (x)\right)}}
\newcommand{\bipert}{\ensuremath{
   \left(\intinf dx' \disorderop (x')\right)}}
\newcommand{\ketirr}{\ensuremath{
   \left| N,M,n\right\rangle}}
\newcommand{\brairr}{\ensuremath{
   \left\langle N,M,n\right|}}
\newcommand{\bra}{\ensuremath{\left\langle}}
\newcommand{\ket}{\ensuremath{\right\rangle}}
\newcommand{\first}{\ensuremath{1^{\mathrm{st}}}}
\newcommand{\second}{\ensuremath{2^{\mathrm{nd}}}}
\newcommand{\third}{\ensuremath{3^{\mathrm{rd}}}}
\newcommand{\dlya}{\ensuremath{\Delta (\log (\frac{y}{a}))}}
\newcommand{\sign}{\ensuremath{\mathrm{sign}}}
\newcommand{\ve}{\varepsilon}

\catcode`\@=11
\@addtoreset{equation}{section}

\begin{document}
\title{{Random Defect Lines in  \\ Conformal Minimal Models}}
\author{\hspace{-0.35in} M. Jeng,   \ \ A. W. W. Ludwig \\
{\it \hspace{-0.6in} Physics Department, University of California, 
Santa Barbara, CA 93106-4030} }
\maketitle

\begin{abstract}
\noindent We analyze the effect of adding quenched disorder along a 
defect line in the 2D conformal minimal models using replicas.
The disorder is realized by a random applied magnetic
field in the Ising model, by fluctuations in the ferromagnetic bond coupling
in the Tricritical Ising model and Tricritical Three-state Potts 
model (the $\phi_{12}$ operator), etc..  We find
that for the Ising model, the defect renormalizes to two decoupled half-planes
without disorder, but that for all other models, the defect renormalizes to a
disorder-dominated fixed point. Its critical properties are studied
with an expansion in $\eps \propto 1/m$ for the $m^{\rm th}$ Virasoro 
minimal model.
The decay exponents $X_N=\frac{N}{2}(1-\frac{9(3N-4)}{4(m+1)^2}+
\mathcal{O}(\frac{3}{m+1})^3)$ of the $N^{\rm th}$ moment of the
two-point function of $\phi_{12}$ along the defect are obtained
to 2-loop order, exhibiting multifractal behavior.
This leads to a typical decay exponent
$X_{\rm typ}=\frac{1}{2} (1+\frac{9}{(m+1)^2}+\mathcal{O}(\frac{3}{m+1})^3)$.
One-point functions are seen to have a non-self-averaging amplitude.
The boundary entropy is larger than that of the pure system by
order $1/m^3$.

As a byproduct of our calculations, we also obtain to 2-loop order
the exponent
$\tilde{X}_N=N(1-\frac{2}{9\pi^2}(3N-4)(q-2)^2+\mathcal{O}(q-2)^3)$
of the $N^{\rm th}$ moment of the energy operator in the
q-state Potts model with {\it bulk} bond disorder.

\end{abstract}

\pagebreak


\section{Introduction}

Conformal symmetry tends to emerge in
pure (homogeneous and rotationally invariant)
 2-D Statistical Mechanics models at their critical points.
This high degree of symmetry severely constrains these
theories, so that these critical points
are well understood.
Many models, including the Ising
model, Tricritical Ising model, and Tricritical Three-state Potts model, are
part of a class of conformal field theories known as Virasoro
minimal models~\cite{CFT.1,CFT.2,CFT.3}. The Ising model~\cite{IM.expt},
Tricritical Ising model~\cite{IM.expt,TIM.expt.1,TIM.expt.2}, and Tricritical
Three-state Potts model~\cite{T3SPM.expt,T3SPM.expt.2} have all been 
realized experimentally  in adsorbed monolayer systems.

Because physical systems always have impurities, it is important to consider 
the effect of quenched disorder on the critical behavior
of these theories. When disorder is 
added to the bulk, random fields are usually relevant, but random bonds
may or may not be (see e.g. Ref.'s~\cite{Cardybook,Harris}). 
It is possible to show, 
for certain systems, 
that with the addition of disorder the system renormalizes 
into an infra-red fixed point. In fact, a  
rigorous theorem shows that when disorder in the order parameter is added
to a system undergoing a first-order phase transition, the latent heat
vanishes~\cite{Aizenmann}, and a $2^{\rm nd}$ order transition
can be expected.

However, one generally does not know
a priori whether the new critical point is disorder-dominated. A number of 
studies have reported cases in which the addition of quenched disorder to
a pure 2-D model at its critical point resulted in another
pure (that is, non-random)
critical model~\cite{purepure.1,purepure.2,purepure.3,purepure.4,
purepure.5,Cardy:FermionRG}. Other studies found
quenched disorder to result in new disorder-dominated fixed 
points~\cite{puredis.1,puredis.2,puredis.3,Ludwig:OPE.RG,JacobsenCardy}. 
One can often  see that a critical point is
disorder-dominated by showing that various universal
quantities are not self-averaging (that measurements 
for a specific `typical' sample may differ 
substantially from those on 
an average one). One particularly interesting manifestation of this is 
multifractal behavior, which occurs when an 
infinite hierarchy of independent scaling dimensions are 
associated with a single 
operator~\cite{multifractal.1,multifractal.2,multifractal.3}.
(See for example~\cite{multifractal.4} for a relevant discussion.)

All the above studies have focused on the effects of adding quenched disorder
to the bulk of a system. However, it is also interesting to consider the case 
where the 2-D model has a defect along which impurities have
clustered. In this paper we consider the effects of adding quenched disorder
only along a defect line, in each of the minimal models;  these models
are labelled by an index $m$, $m\geq 3$. Each $m$ represents a different
model : the Ising model ($m=3$), Tricritical Ising model ($m=4$), 
Tetracritical Ising model ($m=5$),
Tricritical Three-state Potts model ($m=6$), etc. . .
In Section~\ref{sec-model} we introduce our defect model, adding
quenched disorder in the coupling to $\phi_{12}$ (an operator in the Kac 
Table~\cite{CFT.1}) using replicas. The physical meaning of this disorder varies
from model to model, representing a random magnetic field for the Ising
model, but fluctuations in the ferromagnetic coupling (or chemical
potential) in the Tricritical Ising model and Tricritical 
Three-state Potts model. We find that the Ising model renormalizes
to a non-random model (consisting of decoupled half-planes with 
free boundary conditions and no random magnetic field), while all other 
models renormalize to disorder-dominated fixed points. Just as with
bulk disorder, disorder on the defect can result in either a pure or
disorder-dominated fixed point. 

In Section~\ref{sec-RG.D} we calculate the renormalization group
equation of the strength of the disorder  $\Delta$,
 to 2-loop order by minimal subtraction~\cite{Ludwig:OPE.RG,dots:RG}. We
find that arbitrarily weak disorder grows, and flows to a 
new fixed point at which $\Delta$ is of order $1/m$.
This justifies a $1/m$ expansion, 
where critical quantities are calculated perturbatively in $1/m$. The 
"boundary entropy"~\cite{entropy} at the random fixed point is 
found to be $\mathcal{O}(1/m^3)$, and larger than the
entropy of the pure fixed point.

In Section~\ref{sec-scaling} the same scheme is used to find the moments of
correlation functions of the operator $\phi_{12}$ along the 
defect. (Technical details associated
with the irreducible representations of the symmetric 
group~\cite{Ludwig:irreps}
are delegated to
Appendices~\ref{app-RGNM} and~\ref{app-combo}.) It is found 
that the moments fall off as a sum of power laws, the dominant term
decaying as

\begin{eqnarray}
\overline{{\twoptcorrno{1}{2}}^N} \propto\abspow{x_1-x_2}{-2X_N}
\end{eqnarray}

\noindent with

\begin{eqnarray}
\label{eq:mainres}
X_N = \frac{N}{2}\left(1-\frac{1}{4}(3N-4)(\frac{3}{m+1})^2+
	\mathcal{O}(\frac{3}{m+1})^3\right)
\end{eqnarray}

\noindent $X_N<NX_1$, so the operator $\phi_{12}$ exhibits 
multifractal behavior. This leads to a typical decay 
exponent~\cite{Ludwig:irreps} :

\begin{eqnarray}
X_{\rm typical} = \frac{1}{2} \left(1+(\frac{3}{m+1})^2+
\mathcal{O}(\frac{3}{m+1})^3 \right)
\end{eqnarray}

In Section~\ref{sec-OPFmoments} we calculate 
one-point functions of the same operator off the defect line.
The universal (normalized) amplitudes of the moments are 
found to be non-self-averaging, wheras the power law is 
self-averaging.

In Section~\ref{sec-Ising}, the Ising model, which requires special
considerations due to the presence of an additional
marginal (boundary) operator, is 
analyzed. It is argued that when a random magnetic field
is added along the defect line, the system renormalizes to two decoupled 
half-plane Ising models with free boundary conditions and no disorder. 

We finally note that the manipulations needed to get the exponents
in  Eq.(\ref{eq:mainres}) are similar 
to those needed to get the decay exponents of moments of
two-point functions of $\ve$, the energy operator, for the
$q$-state Potts model with {\it bulk} disorder in the bond
strength. This model has been analyzed elsewhere by
expanding about $q=2$ (the Ising 
Model)~\cite{Ludwig:OPE.RG,JacobsenCardy,Ludwig:irreps,dots:RG},
and as a byproduct of our calculations here, we 
find in Section~\ref{sec-scaling}, 
Eq.(\ref{eq:qspm.main}-\ref{eq:qspm.typ}), that

\begin{eqnarray}
	\overline{<\ve (x_1)\ve (x_2)>^N} \propto
	\abspow{x_1-x_2}{-2\tilde{X}_N} 
\end{eqnarray}

\noindent where

\begin{eqnarray}
\tilde{X}_N = N\left(1-\frac{2}{9\pi^2} (3N-4) (q-2)^2+
	\mathcal{O}(q-2)^3\right)
\end{eqnarray}

Results of numerical transfer matrix calculations for the random
bond q-state Potts model can be found in~\cite{JacobsenCardy}.


\section{The Defect Model}
\label{sec-model}

We start with a Virasoro minimal conformal field 
theory~\cite{CFT.1,CFT.2,CFT.3}
labelled by an integer $m$, $m\geq 3$, and perturb it along a defect line. 
The perturbed action is

\begin{eqnarray}
S = S_m+\intinf dx\;h(x)\phi_{12}(x,y=0)
\label{eq:action}
\end{eqnarray}

\noindent $S_m$ is the action of the unperturbed conformal field
theory. $h(x)$ is a random coupling, is picked from a Gaussian 
probability distribution with zero mean and variance $\Delta_0$, and
is uncorrelated along the defect:

\begin{equation}
\overline{h(x)}=0\; , \;\;\;\;
\overline{h(x)h(x')}=2\Delta_0\;\delta(x-x')
\label{eq:probh}
\end{equation}

\noindent The  overbar indicates the disorder average.
The operator $\phi_{12}$ is located at position
$(p,q)=(1,2)$ in the  Kac Table~\cite{CFT.1}, and
exists for any minimal model.
 For the Ising model ($m=3$) it is the spin operator, 
while for the Tricritical Ising
model ($m=4$) and Tricritical Three-State Potts model ($m=6$) it is the
energy operator.  The scaling dimensions $2 h_{pq}$ (twice the
conformal weight) of operators $\phi_{pq}$ in minimal models
are known~\cite{CFT.1} :

\begin{eqnarray}
\label{eq:hpq}
2 h_{pq} & = &\frac{[(m+1)p-mq]^2-1}{2m(m+1)} \ \ ,\\
\nonumber \lefteqn{\hspace{-1.55in} \mathrm{which}\ \mathrm{gives}} \\
\label{eq:h12}
2 h_{12} & = &\frac{1}{2}-\frac{3}{2(m+1)}
\end{eqnarray}

\noindent The replicated~\cite{replica} effective action is

\begin{equation}
S_m^{replica} = \sum_{\alpha=1}^{n} S_m^{\alpha} +
   \Delta_0\intinf dx\disorderop(x,y=0) \ \ ,
\label{eq:fullS}
\end{equation}

\noindent where $n$ is the number of replicas, and we take $n\rightarrow 0$
at the end of the calculation. 

We have ignored higher cumulants of the probability distribution 
of $h(x)$, because power counting shows that they are 
irrelevant\footnote{
For the lower values of $m=4$ or $5$, the $4^{\rm th}$ order cumulants are
respectively relevant and marginal, but the $2^{\rm nd}$ order 
cumulant is more relevant, and presumably will be physically
dominant. While the $4^{\rm th}$ order cumulant is
not irrelevant at the unperturbed fixed point for $m=4$ or $m=5$, 
it is irrelevant for large $m$, and we expect it to be irrelevant 
for $m=4$ or $5$ at the new disordered fixed point. This would 
be similar to the  Wilson-Fisher fixed point,
where the $\phi^6$ operator becomes relevant at the Gaussian fixed 
point below three dimensions, wheras it is in fact irrelevant 
at the new fixed points obtained by an expansion
about four dimensions.}
in the R.G. sense, 
for large $m$, and we will be expanding about large $m$.

We have also dropped the terms with $\alpha=\beta$
in Eq.(\ref{eq:fullS}). The terms with $\alpha=\beta$ produce a 
non-random $\phi_{13}$ by the conformal fusion rules~\cite{CFT.1}.
Upon renormalizing
they will generate other non-random $\phi_{1q}$ with $q\geq 3$. However,
from Eq.(\ref{eq:h12}) we can see that all these terms are irrelevant 
 except when $q=m=3$. So when $m\neq 3$ the perturbation in
Eq.(\ref{eq:fullS}) is the only relevant one. 
The following analysis in Sections~\ref{sec-RG.D}-~\ref{sec-OPFmoments}
will assume this, and will thus
only hold for $m\neq 3$. This will not be too restrictive, since all
quantitites will be calculated by expansion in $\eps = \frac{3}{m+1}$,
and will thus be based on large $m$. But if $m=3$ (the
Ising model), then $\phi_{13}$ (the energy operator) is marginal, and we
need to include the effects of a constant coupling to $\phi_{13}$.
This is done in Section~\ref{sec-Ising}.


\section{Renormalization of  the disorder strength $\Delta$}
\label{sec-RG.D}

\noindent We calculate the renormalization group equation for
$\Delta$ to 2-loop order by minimal 
subtraction~\cite{Ludwig:OPE.RG,dots:RG}.
To calculate the renomalization of $\Delta_0$, we want to know,
given a microscopic disorder strength $\Delta_0$, what effective 
disorder strength $\Delta (r)$ is seen on large length scales $r$.
In a region of size $r$, we expand out
$\mathrm{exp}\{S_{m}^{replica}\}$ in powers of $\Delta_0$.
$(\Delta_0)^p$, with $p\geq 0$, will come with $p$ disorder
operators at various points in the region of size r. 
$p=2,3$ or more disorder operators may look, using repeated
operator product expansions, like a single disorder
operator on larger length scales (or size $r$), and thus create a new
effective disorder strength $\Delta (r)$. This can be represented
schematically as :

\begin{eqnarray}
\nonumber
\lefteqn{\hspace{-4.0in}
\perturbation (x)+\frac{1}{2}\left(\perturbation (x)\right)^2 + } \\
+ \frac{1}{6}\left(\perturbation (x) \right)^3+\ldots  
\longrightarrow \Delta\intinf dx\disorderop (x)
\end{eqnarray}

This analysis will of course generate numerous terms besides the 
disorder operator.  However, as noted above, these terms are all
irrelevant in the RG sense, so we will
not calculate these terms.

For each power of $\Delta_0$, the integrals generated above
are regulated at
short distances by analytic continuation in $\eps\equiv\frac{3}{m+1}$,
and at large distances by an infrared cutoff $r$. The calculation, which
uses the method of~\cite{dots:RG}, is 
done in Appendix~\ref{app-pertD}, where we obtain

\begin{equation}
\label{eq:del(r)}
\Delta(r) = r^{\epsilon}\Delta_0 +
	4(n-2)\frac{r^{2\epsilon}}{\epsilon}\Delta_0^2 
	-4(n-2)\left[2-4(n-2)\frac{1}{\epsilon}\right]
		\frac{r^{3\epsilon}}{\epsilon}
	\Delta_0^3
	+\mathcal{O}(\Delta_0^4)
\end{equation}

\noindent When we calculate the beta function by taking
a derivative with respect to $\log(r)$, we find that
the poles in $\epsilon$ cancel, as they must for any physical
quantity. The result is

\begin{equation}
\label{eq:betafunc}
\beta(\Delta)=\frac{d\Delta}{d(\log (r))}
	= \epsilon\Delta - 8\Delta^2 + 32\Delta^3 +
	  \mathcal{O}(\Delta^4)
\end{equation}

\noindent where we have taken the replica limit
$n\rightarrow 0$ (and have ceased writing the 
$r$ dependence of $\Delta(r)$). The RG flows take the unperturbed
theory to a new infrared fixed point with disorder $\Delta_*$. 
Solving for $\Delta_*$ by putting $\beta(\Delta_*)=0$ gives

\begin{equation} \label{eq:fp}
\Delta_*=\frac{\epsilon}{8}+\frac{\epsilon^2}{16}+
	\mathcal{O}(\epsilon^3),
\qquad \qquad (\epsilon \equiv {3\over m+1})
\end{equation}

\noindent The new fixed point is at a distance
of order $\eps$ from the unperturbed theory,
and so we can calculate physical quantities by expanding in powers of 
$\eps$, which is small for minimal models with large
index $m$. This is analogous
 to the Wilson-Fisher epsilon expansion in $d=4-\eps$ dimensions. 
We will see below that the new fixed point is disorder-dominated.
Because the analysis is based on expansion in powers of
$\eps=\frac{3}{m+1}$, we really only show that there is a new
disorder-dominated fixed point for large $m$ -- but we expect
no qualitative changes for lower $m$, such as $m=4$. However, we
again note that $m=3$, the Ising model, is qualitatively different,
and will be treated separately.

The `boundary entropy' (the universal constant independent of the system
size appearing in the disorder-averaged free energy) associated with 
the defect line,  calculated~ as in Ref.\cite{entropy}, is
found to be

\begin{equation}
\frac{\delta g}{n} = \frac{\pi^2 \eps^3}{96} + \mathcal{O}({\eps^4}).
\end{equation}

\noindent Note that the entropy has increased from that of the
unperturbed system (where it vanishes). 
This is to be contrasted with the case of
a pure (non-random and unitary) system, where
the entropy~\cite{entropy} is expected to only decrease  upon
 renormalization.


\section{Scaling Dimensions of Moments of $\phi_{12}$}
\label{sec-scaling}

\noindent We now look at 
{\scriptsize $\overline{{\twoptcorrno{1}{2}}^N}$}, the 
disorder-averaged $N^{\rm th}$ moment of the  2-point 
function  for points $x_1$ and $x_2$ which lie both
 near the defect and far from each other (Figure 1). 
We want to see how these moments fall off at large distances.
As formulated in replicas we have

\begin{equation}
\label{eq:basicdecay}
\overline{{\twoptcorrno{1}{2}}^N} =
\left< \prod_{\alpha=1}^{N} \phi_{12}^{\alpha}(x_1)
       \prod_{\beta=1}^{N} \phi_{12}^{\beta}(x_2) \right>
\end{equation}

As explained in~\cite{Ludwig:irreps}, we cannot simply calculate
the dimension of $\prod_{\alpha=1}^{N} \phi_{12}^{\alpha}$, because 
this operator is not multiplicatively renormalizable.
 Instead, it is
a sum of independent scaling operators with different scaling 
dimensions. In the limit where the pure and disordered
fixed points collide ($m=\infty$), we have numerous operators with
the same dimension. For example, looking at
$N=2$, $\phi_{12}^1\phi_{12}^2$
and $\phi_{12}^3\phi_{12}^7$ have the same scaling dimension
and are equally good operators (note that the replica limit
$n\rightarrow 0$ is not taken until after all calculations are
completed). In general, for the $N^{\rm th}$ moment, and with 
$n$ replicas, we have $n\choose N$ operators with the same dimension
at $m=\infty$ ; because they all have the same scaling
dimension at $m=\infty$, any linear combination of them also
has this scaling dimension at $m=\infty$.

However, when we move to $m<\infty$, only appropriate linear 
combinations of these operators will have well-defined scaling
dimensions. The entity $\prod_{\alpha=1}^{N} \phi_{12}^{\alpha}$
which appears in Eq.(\ref{eq:basicdecay})
will contain all of these scaling
operators, so that the average 
{\tiny $\overline{{\twoptcorr{1}{2}}^N}$} will decay as a 
mixture of power laws, and
at large distances will be dominated by the scaling operator
with the lowest dimension. We thus need to calculate the dimension
of each of these scaling operators. 
The appropriate multiplicatively renormalizable operators 
transform in irreducible representations of the 
symmetric group~\cite{Ludwig:irreps} :

\begin{equation}
\label{eq:NMnirrep}
\Oirr = \irrsum{\alpha}{M}\irrterm{\alpha}{M}{N}
\end{equation}

\noindent where $0\leq M\leq N$. 

To calculate the scaling dimensions of $\Oirr$, we will  add the term
\linebreak $\Delta_{NMn,0} \intinf \; dx \Oirr(x)$ to 
the action. As in section~\ref{sec-RG.D}, we calculate the renomalization of
$\Delta_{NMN,0}$ to $\Delta_{NMn}(r)=Z_{NMn}(r)\Delta_{NMn,0}$
on length scales of size $r$, by expanding the action in powers
of $\Delta_0$:

\begin{eqnarray}
\nonumber
\lefteqn{\hspace{-3.15in}
\Delta_{NMn,0}\biOirr+} \\
\nonumber
\lefteqn{\hspace{-3.15in}
 + \Delta_{NMn,0}\Delta_0\biOirr\bipert+} \\
\nonumber
\lefteqn{\hspace{-3.15in}
 + \frac{1}{2}\Delta_{NMn,0}\Delta_0^2\biOirr\bipert^2+\ldots }\\
 \longrightarrow Z_{NMn}(r)\Delta_{NMn,0}\biOirr
\end{eqnarray}

In Appendix~\ref{app-RGNM} we check that to 2-loop order we don't need
to worry about mixing with other operators. $Z_{NMn}(r)$ is
calculated with the same type of integrals as in the last section (they
are again regulated at short distances by analytic continuation in
$\eps$ and at large distances by an infrared cutoff $r$). However,
technical combinatorial complexities arise in counting the number
of contractions associated with various irreducible representations
of the symmetric group. They are delegated to
Appendices~\ref{app-RGNM} and~\ref{app-combo}. The result is

\begin{eqnarray}
\nonumber Z_{NMn}(r)=1 & + & 2\OPEb\frac{r^{\epsilon}}{\epsilon}\Delta_0
   -4\left(N(n-N)+(N-1)\OPEb\right)
   \frac{r^{2\epsilon}}{\epsilon}\Delta_0^2 \\
   & + & \left(2(\OPEb)^2+4(n-2)\OPEb\right)
	 \frac{r^{2\epsilon}}{\epsilon^2}\Delta_0^2
   +\mathcal{O}(\Delta_0^3)
\end{eqnarray}

\noindent where we have defined

\begin{equation}
\label{eq:OPEb}
\OPEb\equiv 2\left((N-M)n-N^2+M(M-1)\right)
\end{equation}

\noindent $\OPEb$ is a solely combinatorial factor 
which arises from counting
the number of replica contractions associated with $\Oirr$.
We then use Eq.(\ref{eq:del(r)}) to rewrite the series in
$\Delta_0$ as a series in $\Delta$, yielding $\gamma_{NMn}(\Delta)$.

\begin{eqnarray}
\nonumber\gamma_{NMn}(\Delta) & \equiv &
	\frac{d(\log (Z_{NMn}(r)))}{d(\log (r))} \\
	& = & 2\OPEb\Delta - 8(N(n-N)+(N-1)\OPEb)\Delta^2 +\mathcal{O}(\Delta^3)
\end{eqnarray}

\noindent Note that the poles in $\eps$ again cancel. To
get the scaling dimensions of $\Oirr$ at the disordered fixed point,
we take $n\rightarrow 0$ and $\Delta\rightarrow\Delta_*$, getting

\begin{equation}
\gamma_{NM}(\Delta_*)=\frac{\eps}{4}\OPEbb +
	\frac{\eps^2}{8}(N^2-(N-2)\OPEbb) +
	\mathcal{O}(\eps^3)
\end{equation}

\noindent In the unperturbed theory, for all $M$, 
{\scriptsize $\overline{{<\Oirr(x_1)\Oirr(x_2)>}^N}$} will
fall off as $\mid x_1 - x_2 \mid^{-4Nh_{12}}$.
But with the defect it will fall off as
$\mid x_1 - x_2 \mid^{-2X_{NM}}$, where
$2X_{NM}=4Nh_{12}-2\gamma_{NM}(\Delta_*)$.  Because
$\prod_{\alpha=1}^{N} \phi_{12}^{\alpha}$ is a linear combination of the
scaling operators $\Oirr$, the moment 
{\tiny $\overline{{\twoptcorrno{1}{2}}^N}$} 
will be a sum of terms decaying with powers $X_{NM}$. For 
$\mid x_1 - x_2 \mid$ large, this will be dominated by the smallest
power, and it is easy to see that this is $X_N \equiv X_{NN}$, 
giving our main result :

\begin{eqnarray}
\overline{{\twoptcorrno{1}{2}}^N}\propto\abspow{x_1-x_2}{-2X_N} \\
\lefteqn{\hspace{-2.3in}X_N = \frac{N}{2}(1-\frac{\eps^2}{4}(3N-4)+
	\mathcal{O}(\eps^3))}
\end{eqnarray}

\noindent $X_N<N X_1$, so the system is
not self-averaging. Instead, we have an infinite number of 
independent scaling dimensions all 
associated with
the single operator $\phi_{12}$.
Note that $X_{N_1}/N_1>X_{N_2}/N_2$ for $N_2>N_1$, as
required by convexity.  The result above also yields 
the typical~\cite{Ludwig:irreps} exponent:

\begin{equation}
X_{\mathrm{typical}}=\frac{1}{2}(1+\eps^2+\mathcal{O}(\eps^3))
\end{equation}

The combinatorial problems encountered here are the same
as those for the q-state Potts model with disorder in the {\it bulk}
ferromagnetic couplings. The specific integrals are different
but have already been done in~\cite{dots:RG}.
The only new difficulty is that to two loop order, the 
operators $\Oirr$ are not longer always multiplicatively renormalizable,
but instead mix with other descendent operators. However, luckily,
the leading and subleading operators, $\mathcal{O}_{NNn}$
and $\mathcal{O}_{N,N-1,n}$, remain multiplicatively 
renormalizable (see Appendix~\ref{app-RGNM} for details).
So the combinatorics
in Appendices~\ref{app-RGNM} and~\ref{app-combo} also give the 2-loop 
result for the decay exponent of the $N^{\rm th}$ moment
of the two-point function of the energy operator $\ve$,
in a Potts model with bulk bond disorder. We have:

\begin{eqnarray}
\label{eq:qspm.main}
\lefteqn{\hspace{-2.55in}
	 \overline{<\ve (x_1)\ve (x_2)>^N} \propto
 	 \abspow{x_1-x_2}{-2\tilde{X}_N} }  \\
\tilde{X}_N = N\left(1-\frac{2}{9\pi^2} (3N-4) (q-2)^2+
	\mathcal{O}(q-2)^3\right)
\end{eqnarray}

\noindent This also gives us the typical decay exponent :

\begin{equation}
\label{eq:qspm.typ}
\tilde{X}_{\mathrm{typical}}=1+\frac{8}{9\pi^2}(q-2)^2+\mathcal{O}(q-2)^3
\end{equation}


\section{Moments of the One-point Function of $\phi_{12}$ }
\label{sec-OPFmoments}

\noindent We now calcuate the disorder-averaged moments of the 
one-point function of $\phi_{12}$, evaluated at a distance $y$ from
the defect line (Figure 2).
The method is similar to that in~\cite{Ludwig:opf}. We no
longer need to worry about the various irreducible 
representations of the symmetric group,
because the one point function of $\Oirr$ vanishes
by symmetry except when $M=0$. The disorder-averaged odd moments
of $\phi_{12}$ vanish because the disorder-averaged system is
symmetric under $\phi_{12}\rightarrow -\phi_{12}$.
The even moments are calculated perturbatively :

\begin{eqnarray}
\nonumber
\lefteqn{ \overline{\bra\phi_{12}(x=0,y)\ket^{2N}} =
\bra\prod_{\alpha =1}^{2N} \phi_{12}^{\alpha} (0,y)\ket }  \\
\nonumber
 & = &  \bra\left[\opprod\right] e^{\perturbation}\ket_{0,\mathrm{cutoff\ a}} \\
\nonumber
 & = & y^{-(1-\eps)N} \sum_{i=N}^{\infty} \frac{1}{i!} (\Delta_0 y^\eps)^i
   \bra \left[\opproda\right] \left[\prod_{j=1}^{i}\intinf dx_j\right] 
   \right. \\
\nonumber
 & & \left. \left[\prod_{j=1}^{i} 
 \left(\disopbb{\beta_j}{\gamma_j}{x_j}\right) \right] 
 \left[\prod_{(k<l)=1}^i 
 \theta\left(\abspow{x_i-x_j}{}-\frac{a}{y}\right)\right]\ket_0 \\
\label{eq:pertOPF}
 & \equiv & y^{-(1-\eps)N} \sum_{i=M}^{\infty} \frac{1}{i!} (\Delta_0 y^\eps)^i
   I_{i}^{(N)}(\frac{a}{y})
\end{eqnarray}

\noindent In the $\second$ line we have introduced a cutoff $a$ to 
regulate the short-range 
\linebreak divergences. In the $\third$ line we have expanded out the
effect of the defect perturbatively in $\Delta_0$, and used the conformal
symmetry to rescale each 
\linebreak expectation value by $y$. In the lowest order 
term, $I_N^{(N)}$,
to get a nonzero expectation value, the 2N operators on the defect must 
lie in different replicas -- so nothing special happens when two 
defect terms get close together, and we can drop the cutoff $a$.

\begin{equation}
I_N^{(N)}(\frac{a}{y}) = (2N)! 
\left[\intinf \;\frac{dx}{(x^2+1)^{1-\eps}}\right]^N
=(2N)! \pi^N (1+\mathcal{O}(\frac{a}{y},\eps))
\end{equation}
  
\noindent We can now use the operator product expansion (OPE)

\begin{eqnarray}
\nonumber
\lefteqn{ \hspace{-0.6in} \left[\disopb{\alpha}{\beta}{x+\delta}\right]
\left[ \disopb{\alpha'}{\beta'}{x}\right] } \\
 & & \longrightarrow\frac{4(n-2)}{\delta^{1-\eps}}\disopb{\alpha}{\beta}{x}
\ \ \ \mathrm{for}\ \ \ \delta\rightarrow 0 
\end{eqnarray}

\noindent to get the leading poles in $\eps$ for the higher-order terms, by
taking derivatives of $I_{i}^{(N)}(\frac{a}{y})$ with respect to $\frac{a}{y}$.
When we take derivatives of the step function
$\theta(\abspow{x_i-x_j}{}-\frac{a}{y})$,
we bring two operators close together, and so can use the OPE. We get

\begin{equation}
\label{eq:OPEiter}
\frac{\partial I_{i}^{(N)}(\frac{a}{y})}{\partial(a/y)} =
-4i(i-1)(n-2)(\frac{a}{y})^{-1+\eps} I_{i-1}^{(N)}(\frac{a}{y})\ \ \ \mathrm{for}
\ \ \ i>N
\end{equation}

\noindent We already have $I_N^{(N)}$, so we can now
repeatedly integrate to get $I_i^{(N)}$ for $i>N$.
$I_i^{(N)}$ must be finite as $\eps\rightarrow 0$ for $a\neq 0$,
because with a finite cutoff $a$, there are no ultraviolet
singularities in the integrals for $I_i^{(N)}$ -- this 
requirement determines all the constants of integration. So, for
example, the leading term in $I_{N+1}^{(N)}(\frac{a}{y})$ is

\begin{eqnarray}
\nonumber I_{N+1}^{(N)}(\frac{a}{y}) & = &
-4N(N+1)(n-2)I_N^{(N)}(0)\left[\frac{1}{\eps}(\frac{a}{y})^\eps+
			\mathrm{constant}\right]
(1+\mathcal{O}(\frac{a}{y})) \\
\nonumber & = & -4N(N+1)(n-2)I_N^{(N)}(0)
   \left(\frac{(\frac{a}{y})^\eps -1}{\eps}\right)
   (1+\mathcal{O}(\frac{a}{y})) \\
 & \stackrel{\eps\rightarrow 0}{\longrightarrow} &
-4N(N+1)(n-2)I_N^{(N)}(0) \log (\frac{a}{y})
\end{eqnarray}

\noindent (Note that this does diverge as $a\rightarrow 0$.) 
More generally, by repeatedly integrating 
Eq.(\ref{eq:OPEiter}) and using the requirement that all terms be 
finite as $\eps\rightarrow 0$ for $a\neq 0$, we find

\begin{equation}
I_i^{(N)}(\frac{a}{y})=\frac{i!(i-1)!}{(i-N)!N!(N-1)!}
   \left( -4(n-2)\log (\frac{a}{y})\right)^{i-N} I_N^{(N)}(0),\;\;\; i\geq N
\end{equation}

\noindent Now taking the leading divergences of each term in
Eq.(\ref{eq:pertOPF}) gives

\begin{eqnarray}
\nonumber
\overline{\bra\phi_{12}(y)\ket^{2N}} & = & 
	\frac{(\Delta_0 y^{-1+2\eps})^N I_N^{(N)}(0)}{N!(N-1)!}
	\sum_{i=N}^{\infty} \frac{(i-1)!}{(i-N)!}
	(-4(n-2)\log (\frac{a}{y})y^\eps \Delta_0)^{i-N} \\
\label{eq:opfsum}
     & = & \frac{I_N^{(N)}(0)}{N!} y^{-(1-\eps)N}
       \left( \frac{\Delta_0 y^\eps}{1+4(n-2)\log (\frac{a}{y}) 
	y^\eps \Delta_0} \right)^N
\end{eqnarray}

In the front we have extracted a constant and the
expected power-law 
dependence $y^{-(1-\eps)N}=y^{-2h_{12}N}$.
The remaining terms are written as a function $F[a,y,\Delta_0]$
of the large distance $y$ and the microscopic variables $a$ and
$\Delta_0$. However, we know from the Callen-Symanzik equation
that this can be rewritten in terms of a single scaling function 
dependent only on the renomalized coupling for length scales of 
order $y$, $\dlya$~\cite{Ludwig:opf}. Explicitly,

\begin{equation}
F[a,y,\Delta_0] = F[ae^\ell,y,\Delta(\ell)] = F[y,y,\dlya]
 = G[\dlya]
\end{equation}

\noindent We rewrite $\Delta_0$ in terms of $\dlya$ by integrating
Eq.(\ref{eq:betafunc}) to $\first$ order, getting

\begin{equation}
\Delta_0 y^\eps = \dlya + 8\log (\frac{y}{a}) (\dlya )^2 +
\mathcal{O}(\dlya)^3 + \ldots
\end{equation}

\noindent Substituting this into Eq.(\ref{eq:opfsum}), we 
indeed get that the leading divergences at all orders of 
perturbation theory sum up to give a function which depends only
on $\dlya$ :

\begin{equation}
\overline{{\bra \phi_{12}(y) \ket}^{2N}} = 
\frac{(2N)!}{N!} \left(\frac{\pi\dlya}{y^{2h_{12}}}\right)^N \ \ , \ \ \ \ 
\overline{{\bra \phi_{12}(y) \ket}^{2N+1}}=0
\end{equation}

We see that the amplitudes of one-point functions are not self-averaging :
$\overline{{\bra \phi_{12}(y) \ket}^{2N}} \neq
[\overline{{\bra \phi_{12}(y) \ket}^{2}}]^N$. That
is, the average of the $N^{\rm th}$ power is different than the
$N^{\rm th}$ power of the average. The amplitude ratios are
universal properties of the random defect fixed point. 

Also note that 
$\frac{1}{N_1} \log \left[ \; \overline{{\bra \phi_{12}(y) \ket}^{2N_1}} \; \right] <
 \frac{1}{N_2} \log \left[ \; \overline{{\bra \phi_{12}(y) \ket}^{2N_2}} \; \right] $
for $N_1<N_2$, as required by convexity. 


\section{Ising Model}
\label{sec-Ising}

We now look at the defect introduced in Eq.(\ref{eq:action})
for the special case of the Ising model.  As noted at the end of 
Section~\ref{sec-model}, the analysis in the three sections above 
fails when $m=3$, because along with the perturbation in 
Eq.(\ref{eq:fullS}), we generate the marginal 
non-random operator $\phi_{13}$. The minimal model with
$m=3$ corresponds to the Ising model,
and in this model $\phi_{12}$ is the spin operator ($\sigma$) and 
$\phi_{13}$ is the energy operator ($\ve$).
So in this case our defect consists of 
a random applied magnetic field and a non-random energy operator.
Our action is

\begin{equation}
\label{eq:IMaction}
S=S_{Ising}+ \intinf dx \left[ \Delta\sum_{\alpha\neq\beta}^n
\sigma_{\alpha}(x,0)\sigma_{\beta}(x,0)+
\lambda\sum_{\alpha=1}^{n}\ve_{\alpha}(x,0) \right]
\end{equation}

In the lattice formulation of the Ising model, perturbing with 
the energy operator is the same as changing the bond strength.
Defects where the bond strength is changed along a single line
have been studied and solved exactly 
by Bariev and McCoy et. al.~\cite{Bariev,McCoy}.
They dealt with cases where the bond 
strength was changed only in the bonds perpendicular to the 
defect (the ladder geometry -- Figure 3), or only in the bonds parallel to 
the defect (the chain geometry -- Figure 4). 
Our defect, for the Ising model, is thus a perturbation with a
random magnetic field, of these exactly solved defects.
The scaling dimension of the spin operator along the 
defect is $x_\sigma =\frac{1}{2}g(\lambda)^2$, where

\begin{eqnarray}
\label{eq:glambdaladder}
g(\lambda) & = & \frac{2}{\pi}\tan^{-1}
\left(\frac{\sqrt{2}-1}{\tanh(K_c+\lambda)}\right)\ \ \ 
\mathrm{for\ the\ ladder\ geometry,} \\
\nonumber \lefteqn{\hspace{-0.7in} \mathrm{and}} \\
\label{eq:glambdachain}
g(\lambda) & = & \frac{1}{\pi}\cos^{-1}
\left(\tanh(2\lambda)\right)\ \ \ 
\mathrm{for\ the\ chain\ geometry}
\end{eqnarray}

Here we use the results of Bariev and McCoy et. al.~\cite{Bariev, McCoy}, 
and along the defect have 
changed the bond coupling from $K_c=\log (\frac{1}{2}(1+\sqrt{2}))$
to $K_c +\lambda$. This is not quite correct, because $\lambda$ is the
coefficient of the energy operator in the continuum formulation, while
$K=K_c +\lambda$ is the coupling in the lattice formulation. However, 
taking this 
difference into account will only give a ($\lambda$-dependent)
rescaling of our renormalizaion group flows, which will not 
affect the qualitative results. 
Some subtleties regarding the branch of the arctangent in
Eq.(\ref{eq:glambdaladder}) for antiferromagnetic ladder
couplings are dealt with in 
Appendix~\ref{app-IsingLadder}. We get the
OPE coefficient $b_{\sigma\sigma\ve}=-g(\lambda)$
from~\cite{KadanoffIs, Kadanoff} : 

\begin{equation}
\sigma(x-\frac{\delta}{2})\sigma(x+\frac{\delta}{2}) =
\delta^{-g^2}[1+\delta g\ve+\ldots]
\end{equation}

\noindent We can now get the
renormalization group equations to 1-loop order solely
from the OPE's~\cite{Ludwig:OPE.RG,Ludwig:opf} : 

\begin{eqnarray}
\label{eq:IMflow1}
\frac{d\Delta}{d\ell}  & = & (1-g(\lambda)^2)\Delta
	-8\Delta^2 \\
\label{eq:IMflow2}
\frac{d\lambda}{d\ell} & = & -4 g(\lambda) \Delta^2
\end{eqnarray}

The flows for the ladder and chain cases are shown in Figures 5 and 6.
The flows for the ladder case show 
that perturbations about the point $\Delta=\lambda=0$ (the defect-free 
point) eventually
flow to the point with $(\Delta,\lambda)=(0,-K_c)$.
This value of $\lambda$ corresponds to vanishing bond strength 
along the ladder. So the renormalization group flow
takes us to a point with
two decoupled half-plane Ising models and
no random magnetic field. 

We can check our result by looking at the flows around
the decoupled point. At this point the spins can be represented by
free fermions, and we can calculate the renormalization group 
equations about this point. These results 
agree with the flows around the decoupled point obtained from
Eq.(\ref{eq:IMflow1}) and Eq.(\ref{eq:IMflow2}). In making the
comparison, it is important to note that in the replica 
formalism, disorder in the magnetic field corresponds to an operator
\linebreak $\sum_{\alpha=1}^n\sum_{\beta=1}^n
\sigma_{\alpha}(x,0)\sigma_{\beta}(x,0)$,
but in Eq.(\ref{eq:IMaction}), $\Delta$ is the coefficient of 
this operator
with the $\alpha=\beta$ terms removed. This means that $\Delta$ is not really
the strength of the random magnetic field, but a linear combination of
the strength of the random magnetic field and the bond
stength ($\lambda$). So, in figure 5, adding a random magnetic
field to the decoupled point, $(\Delta,\lambda)=(0,-K_c)$, moves
us to a point with $\Delta>0$ and $\lambda>-K_c$, and the 
renormalization group flows from this new point eventually go 
back to the original decoupled point.

The flows in the chain case take perturbations about
the defect-free point to
$(\Delta,\lambda)=(0,-\infty)$. Again, the random
magnetic field has vanished at the new fixed point. To interpret 
this value of $\lambda$,
we look at the spin operator. The dimension of the spin operator 
at $\lambda=-\infty$ is
given by Eq.(\ref{eq:glambdachain}) as
$x_\sigma=\frac{1}{2}g(-\infty)^2=\frac{1}{2}$, which is
the same as the dimension of the spin operator along the 
edge of an Ising model with free boundary conditions. 
The possible boundary conditions of the Ising model have been
completely classified~\cite{Affleck} and the only boundary
condition where the spin operator has dimension 1/2 is the 
free boundary condition. 
This makes sense physically, because $\lambda=-\infty$ 
corresponds to an infinitely
antiferromagnetic coupling that produces alternating spins along
the defect, which upon coarse-graining gives net magnetization zero
everywhere along the line.
We conclude that in the chain case we also flow
to two decoupled Ising models with free boundary conditions. 

If is not hard to see that adding higher cumulant terms of the
magnetic field to Eq.(\ref{eq:IMaction}) will not change
the qualitative results of our 1-loop renormalization group calculations.

So far, we have represented the perturbation in the energy operator
as changing the bond strength either in the vertical or the horizontal
direction, but not in both. More generally, we should represent
the perturbation as changing bond strengths in both directions. 
However, analagously to the ladder and chain cases, we expect that
a more isotropic treatment would only give a different monotonically
decreasing function $g(\lambda)$, and that the point with
\linebreak $(\Delta,\lambda)=(0,g^{-1}(1))$ would still be a stable fixed point
with a large basin of attraction (including the defect-free
model at $\Delta=\lambda=0$). And by the classification of 
Ising model boundary states in~\cite{Affleck},
the point with $(\Delta,\lambda)=(0,g^{-1}(1))$ will always correspond to
two decoupled half-plane Ising models with free boundary conditions
and no random magnetic field.


\section{Conclusions}
\label{conclusions}

We have found a new universality class of disordered 
defect lines. The
defect lines exist in various two-dimensional Statistical
Mechanical models, such as the Tricritical Ising model
and Tricritical 3-state Potts model. The large-distance behavior
of these defect lines has been shown to be disorder-dominated.
Two-point functions along the defect exhibit multifractal
behavior, and universal (normalized) amplitudes of one-point
functions are non-self-averaging. We have also argued that when
a random magnetic field is applied along a single line of the
Ising model, it causes the two sides of the defect to decouple, and
to turn into two half-plane Ising models with free boundary 
conditions and no disorder. 

Results for the defect line in the $m^{\rm th}$ Virasoro minimal
model were obtained by a $1/m$ expansion. However, the physically
most interesting models are at low $m$. For example, $m=4$ corresponds
to the Tricritical Ising model, with a random bond strength (or 
a random chemical potential) along a line. Our calculations show 
that this results in disorder-dominated long-distance behavior.
It would be interesting to understand this model in a more
fundamental and non-perturbative fashion.
The random boundary/defect fixed points
that we have found in this paper are, besides  the bulk random 
bond $q$-state Potts models, a rare case where detailed
analytic information about random critical behavior
is available.
In particular, it would be most interesting to
compare our results for the random defect lines
in minimal models with future  numerical results, such
as those obtained by Jacobsen and Cardy for the
bulk random Potts models\cite{JacobsenCardy}.
 
As a byproduct of our calculations for the defect line, 
we have also obtained multifractal energy-energy correlations 
in the bulk random bond q-state Potts model.

Finally,  we comment on the R.G. analysis
performed under the assumption of
broken replica symmetry.
Such a calculation was done
in~\cite{RSB.theory} for the q-state Potts
model with bulk disorder in the bond strength,
where it was found that 
the replica symmetric disordered fixed
point is unstable to a new fixed point with broken replica 
symmetry. However, numerical tests show that the Potts
model with random bonds is best described by the replica
symmetric fixed point~\cite{RSB.expt.1,RSB.expt.2}.
An  identical analysis to that of~\cite{RSB.theory}  for
the random defect problem that is the subject of the present
paper shows that, again,  the fixed
point considered in this paper is unstable,
and flows to a new stable fixed point with broken replica
symmetry. Thus, our random defect  problem may provide
further insights into the significance of the
replica broken fixed point.

\pagebreak

\noindent {\Large\bf Acknowledgements}

\bigskip

We thank Tom Davis for useful discussions.
This work was supported in part by a Regents Fellowship of the
University of California (M.J.), and in part by the A.P. Sloan
Foundation (A.W.W.L.).


\newpage
\appendix

\section{Renormalization of $\Delta$}
\label{app-pertD}

\noindent We calculate the renormalization of $\Delta$ to second
order. The structure of the calculations closely parallels that in
~\cite{dots:RG}. We get the coefficients in

\begin{equation}
\label{eq:unk.templ}
\Delta(r)=r^{\epsilon}\left(\Delta_0+A_2(r,\epsilon)\Delta_0^2+
	(A_{31}(r,\epsilon)+A_{32}(r,\epsilon) +
		A_{33}(r,\epsilon))\Delta_0^3+\ldots\right)
\end{equation}

\noindent where, as explained below, the different terms come from the 
different types of contractions at each order of perturbation theory.
All integrals are regularized at short distances by analytic
continuation in $\eps\equiv\frac{3}{m+1}$ and at large
distances by an infrared cutoff $r$.


\subsection{First Order}

To lowest order we bring two perturbation terms together

\begin{equation}
\frac{\Delta_0^2}{2} \sepri{1}{2} dx_1 dx_2
   \disop{\alpha}{\beta}{1}\disop{\gamma}{\delta}{2}
\end{equation}

\noindent We get the same perturbation back again (and thus a 
contribution to $\Delta$) when $\beta=\gamma$ and $\alpha\neq\delta$. 
This gives us the $A_2$ term in Eq.(\ref{eq:unk.templ})

\begin{equation}
\label{eq:A2.int}
A_2=2(n-2) \sepri{1}{2} dx_2 \twoptcorr{1}{2}
   = 4(n-2)\frac{r^{\epsilon}}{\epsilon}
\end{equation}

\noindent The $0$ subscript on the correlator indicates
that it is calculated in the defect-free theory.


\subsection{Second Order}
\label{sec-secondRG}

To second order we have

\begin{eqnarray}
\label{eq:genericsecond}
\nonumber\frac{\Delta_0^3}{3!} \sepri{1}{2}\ \ \ \sepri{1}{3} dx_1 dx_2 dx_3
   & & \disop{\alpha}{\beta}{1}\disop{\gamma}{\delta}{2} \\
   & & \disop{\mu}{\nu}{3}
\end{eqnarray}

\noindent This gives us several possible contractions. We get one 
possible contraction when $\beta=\gamma$, $\delta=\mu$, 
$\alpha\neq\delta$, $\beta\neq\nu$, $\alpha\neq\nu$:

\begin{eqnarray}
\label{eq:A31}
A_{31} =  4(n-2)(n-3)\I{1},
\end{eqnarray}

\noindent where we have defined

\begin{eqnarray}
\label{eq:I1}
\nonumber\I{1} & \equiv & \sepri{2}{1} dx_2\sepri{3}{1} dx_3 
   \twoptcorr{1}{2}\twoptcorr{2}{3} \\
\nonumber & = & \sepri{2}{1} dx_2\sepri{3}{1} dx_3
   \twoptcorrb{1}{2}\twoptcorrb{2}{3} \\
 & = & 2r^{2\epsilon}
       \int_{0}^{1}dy\abspow{y}{-1+2\epsilon}
       \int_{\mid z \mid < 1/y} dz \abspow{z}{-1+\epsilon}
		\abspow{1-z}{-1+\epsilon}
\end{eqnarray}

\noindent We have transformed to coordinates
$y=\frac{x_2-x_1}{r}$ and $z=\frac{x_3-x_1}{x_2-x_1}$.
We can now extend the integral over $z$ to go from $-\infty$
to $\infty$, since this will only change $\I{1}$ to $\mathcal{O}(\eps^0)$.
This gives us two integrals which we can do
exactly :

\begin{eqnarray}
\nonumber \I{1} & = & 2r^{2\epsilon}
	\left(\frac{1}{2\epsilon}\right)
	\left(\frac{2^{1-2\epsilon}}{\sqrt{\pi}}\right)
	(1+\cos{\pi\epsilon})
	\Gamma(\epsilon)\Gamma(\frac{1}{2}-\epsilon) 
	+\mathcal{O}(\epsilon^0) \\
\label{eq:I1found}
 & = & 4\frac{r^{2\epsilon}}{\epsilon^2}
	+\mathcal{O}(\epsilon^0)
\end{eqnarray}

\noindent Note that in the integrals we have picked out one point $x_1$ as
special, and integrated over $x_2$ and $x_3$, but in the combinatorial
factor $4(n-2)(n-3)$ we have treated all points symmetrically (i.e., have
treated $\alpha=\delta$, $\beta=\mu$ the same as $\beta=\gamma$,
$\delta=\mu$). This is permissible to this order in perturbation theory, since
we are only concerned with the poles in $\eps$, which result from short range
divergences and are the same for any permutation of the contractions.
We can see this explicitly by considering a different arrangements of the
same contractions:

\begin{eqnarray}
\nonumber
\mathcal{I'}_{1} & \equiv & \sepri{2}{1} dx_2 \sepri{3}{1} dx_3
   \twoptcorr{1}{2}\twoptcorr{1}{3} \\
\label{eq:I1.alt}
 & = & 4\frac{r^{2\eps}}{\eps^2} 
   = \I{1} + \mathcal{O}(\eps^0)
\end{eqnarray}

\noindent We get a second possible set of contractions when
$\alpha=\gamma=\mu$ and $\delta=\nu\neq\beta$:

\begin{eqnarray}
\label{eq:A32}
A_{32} = 4(n-2)\mathcal{I}_2\ ,
\end{eqnarray}

\noindent where we have defined

\begin{eqnarray}
\nonumber
\I{2} & = & \sepri{2}{1}dx_2\sepri{3}{1} dx_3 \left[
   <\phi_{12}(x_1)\phi_{12}(x_2)\phi_{12}(x_3)\phi_{12}(\infty)>_0 \times 
   \right. \\
\label{eq:I2}
   & & \left.\hspace{0.5in} <\phi_{12}(x_2)\phi_{12}(x_3)>_0 - 
   (<\phi_{12}(x_2)\phi_{12}(x_3)>_0)^2\right]
\end{eqnarray}

\noindent The four-point function 
gives the coefficient of three $\phi_{12}$'s
projecting to a single $\phi_{12}$.
We have subtracted off
$(<\phi_{12}(x_2)\phi_{12}(x_3)>_0)^2$ -- this term corresponds to
the contribution from two perturbation
terms (disorder 
operators) getting close, and does not affect the
dimension of the operator at $x_1$. This term is only a contribution
to the free energy, so subtracting it off simply corresponds to 
normalizing the correlation functions. We get the four-point function
from the Coulomb gas formalism~\cite{dots:Coulomb, dots:math}:

\begin{eqnarray}
\nonumber
\lefteqn{\hspace{-0.25in} <\phi_{12}(0)\phi_{12}(1)\phi_{12}(z)\phi_{12}(\infty)>_0 =
\abspow{z}{-1+\eps}\abspow{1-z}{1-\eps /3} } \\
\label{eq:4pf}
 & & \hspace{-0.35in} \left[ {\frac{3k_1}{4} \! \abspow{z}{2-4\eps /3}
 \mid \! F(1-\frac{\eps}{3},2-\eps;2-\frac{2\eps}{3};z) \! \mid^2 \! +
 \! \mid \! F(1-\frac{\eps}{3},\frac{\eps}{3};\frac{2\eps}{3};z) \! \mid^2} \right] ,
\end{eqnarray}

\noindent where we have defined

\begin{eqnarray}
k_1\equiv - \frac{ 4(\Gamma(\frac{2\eps}{3}))^2 \Gamma(2-\eps)
			\Gamma(1-\frac{\eps}{3}) }
		{ 3\Gamma(\frac{\eps}{3}) \Gamma(-1+\eps)
			\Gamma(2-\frac{2\eps}{3}) }
 = 1+\mathcal{O}(\eps)
\end{eqnarray}

\noindent and F is the hypergeometric function. If we take the limit
as $\eps\rightarrow 0$ at fixed $z$ for the four-point function,
we get

\begin{eqnarray}
\nonumber
\lefteqn{ <\phi_{12}(0)\phi_{12}(1)\phi_{12}(z)\phi_{12}(\infty)>_0 
\stackrel{\eps\rightarrow 0}{\longrightarrow} } \\
\label{eq:4pf.e0}
 & & \sign (z)\sign (z-1) + \frac{\sign (z)}{\abspow{z-1}{}} - 
\frac{\sign (z-1)}{\abspow{z}{}}
\end{eqnarray}

We now go back to $\mathcal{I}_2$, which we can calculate by
using the symmetries $x_2 \rightarrow -x_2$ and 
$x_2 \leftrightarrow x_3$ to cut the integration region down by a
fourth, and then
transforming to new coordinates
$y=\frac{x_2-x_1}{r}$ and $z=\frac{x_3-x_1}{x_2-x_1}$.
The integral now exactly factorizes into two one-dimensional
integrals :

\begin{eqnarray}
\nonumber
\I{2} & = & 4r^{2\eps} \int_0^1 dy\ y^{-1+2\eps}
   \int_{-1}^1 dz \left[
  <\phi_{12}(0)\phi_{12}(1)\phi_{12}(z)\phi_{12}(\infty)>_0 \times 
   \right. \\
 & & \hspace{.95in} \left. <\phi_{12}(1)\phi_{12}(z)>_0 - 
 (<\phi_{12}(1)\phi_{12}(z)>_0)^2 \right]
\end{eqnarray}

The y-integral gives $\frac{1}{2\epsilon}$, while the z-integral can 
be rewritten as

\begin{eqnarray}
\nonumber
\int_{-1}^{1} dz & & \hspace{-0.26in} \left[ \rule{0in}{2.8ex}
  <\phi_{12}(0)\phi_{12}(1)\phi_{12}(z)\phi_{12}(\infty)>_0
  <\phi_{12}(1)\phi_{12}(z)>_0 \right. \\
\nonumber
 & & \hspace{0.00in} \left. - <\phi_{12}(0)\phi_{12}(z)>_0
      - \left( <\phi_{12}(1)\phi_{12}(z)>_0 \right)^2 \rule{0in}{2.8ex} \right] \\
\label{eq:z.integral}
 & & \hspace{-0.20in} + \int_{-1}^{1} dz
<\phi_{12}(0)\phi_{12}(z)>_0
\end{eqnarray}

The second integral is exactly $\frac{2}{\epsilon}$. 
It is straightforward to use Eq.(\ref{eq:4pf}) to check that
as $z\rightarrow 1$, the four-point function goes to
$\mid z-1 \mid^{-1+\eps} + \mathcal{O}(z-1)$. So the first
integral converges everywhere, and we can get it's value to
$\mathcal{O}(\eps^0)$ by simply replacing $\eps$ with zero
inside the integral:

\begin{eqnarray}
\nonumber
\int_{-1}^1 dz & & \hspace{-0.32in} \left[ \frac{1}{\abspow{z-1}{}}
   ( \sign (z)\sign (z-1) + \frac{\sign (z)}{\abspow{z-1}{}} - 
     \frac{\sign (z-1)}{\abspow{z}{}} ) \right. \\
\nonumber
 & & \hspace{0.0in} \left. - \frac{1}{\abspow{z}{}} - \frac{1}{\abspow{z-1}{2}} \right]
   + \mathcal{O}(\eps) \\
 & = & -1+\mathcal{O}(\eps)
\end{eqnarray}

\noindent Putting this all together gives

\begin{eqnarray}
\nonumber
\I{2} & = & 4 r^{2\eps} \left(\frac{1}{2\eps}\right)
   \left[ \frac{2}{\eps} -1 + \mathcal{O}(\eps) \right] \\
\label{eq:I2found}
 & = & -2r^{2\eps}\left(\frac{1}{\eps}-\frac{2}{\eps^2}\right)
\end{eqnarray}

\noindent Again, as with $A_{31}$, other possible permutations of this 
contraction like

\begin{eqnarray}
\nonumber 
\I{2}' & = & \sepri{2}{1}dx_2\sepri{3}{1} dx_3
   <\phi_{12}(x_1)\phi_{12}(x_2)\phi_{12}(x_3)\phi_{12}(\infty)>_0 \\
\label{eq:I2.alt}
   & & \hspace{1.37in}<\phi_{12}(x_1)\phi_{12}(x_2)>_0
\end{eqnarray}

\noindent give the same result up to terms of $\mathcal{O}(\eps^0)$. 

Finally, the last possible contraction in Eq.(\ref{eq:genericsecond})
comes from $\alpha=\gamma=\mu$ and $\beta=\delta=\nu$. This is

\begin{eqnarray}
\nonumber
A_{33} & = & \frac{4}{3} \sepri{2}{1} dx_2 \sepri{3}{1} dx_3
   \left[ \left(<\phi_{12}(x_1)\phi_{12}(x_2)\phi_{12}(x_3)\phi_{12}(\infty)>_0
	\right)^2 \right. \\
\nonumber
 & & \hspace{1.6in} -\left.\left( < \phi_{12}(x_2)\phi_{12}(x_3) >_0 
	\right)^2 \right]  \\
\nonumber
 & = & \frac{16}{3} r^{2\eps} \int_0^1 \frac{dy}{y^{1-2\eps}}
     \int_{-1}^{1} dz \left[
   \left(<\phi_{12}(0)\phi_{12}(1)\phi_{12}(z)\phi_{12}(\infty)>_0
	\right)^2 \right. \\
 & & \hspace{1.6in} -\left.\left( < \phi_{12}(1)\phi_{12}(z) >_0 
	\right)^2 \right]
\end{eqnarray}

\noindent The $y$-integral gives $\frac{2}{\eps}$.
The $z$-integral is evaluated similarly to Eq.(\ref{eq:z.integral}).
We rewrite it as

\begin{eqnarray}
\nonumber
\int_{-1}^1 dz & & \hspace{-0.26in} \left[ 
   \left(<\phi_{12}(0)\phi_{12}(1)\phi_{12}(z)\phi_{12}(\infty)>_0
	\right)^2 \right.\\
\nonumber
   & & \hspace{-0.16in} - \left.\left( < \phi_{12}(0)\phi_{12}(z) >_0 \right)^2 
   - \left( < \phi_{12}(1)\phi_{12}(z) >_0 \right)^2 - 1 \right] +\\
\int_{-1}^1 dz & & \hspace{-0.26in} \left[ 1 +
   \left( < \phi_{12}(0)\phi_{12}(z) >_0 \right)^2 \right]
\end{eqnarray}

The second integral can be evaluated exactly, and is
$\mathcal{O}(\epsilon)$. The first integral is nowhere
divergent, so we can get its value to
$\mathcal{O}(\eps^0)$ by taking $\eps$ to 0 inside the integral. Then
squaring Eq.(\ref{eq:4pf.e0}) gives

\begin{eqnarray}
\nonumber
\lefteqn{\hspace{-1.7in}
\left( <\phi_{12}(0)\phi_{12}(1)\phi_{12}(z)\phi_{12}(\infty)>_0 \right)^2 }\\
\nonumber
\lefteqn{\hspace{-1.25in}
\stackrel{\eps\rightarrow 0}{\longrightarrow}
\left(\sign (z)\sign (z-1) + \frac{\sign (z)}{\abspow{z-1}{}} - 
\frac{\sign (z-1)}{\abspow{z}{}} \right)^2 }\\
\lefteqn{ \hspace{-1.25in}
= 1+\frac{1}{z^2}+\frac{1}{(z-1)^2} }
\end{eqnarray}

So both $z$-integrals are $\mathcal{O}(\eps)$, and $A_{33}$ is 
$\mathcal{O}(\eps^0)$. Since $A_{33}$ has no pole in $\eps$,
it can be dropped to this order.
Combining all these results gives Eq.(\ref{eq:del(r)}).


\newpage
\section{Renormalization of $\Delta_{NMn}$}
\label{app-RGNM}

To get the dimensions of the moments of $\phi_{12}$, we need
the coefficients of $\Delta_0$ and $\Delta_0^2$ 
in $Z_{NMn}(r)$, where $\Delta_{NMn}(r)=Z_{NMn}(r)\Delta_{NMn,0}$.
The integrals in this section will be the same as 
the integrals in appendix~\ref{app-pertD}. The only difference 
will be in the combinatorial factors which precede them. In this
section we will use $\brairr$ and $\ketirr$ to represent the bra
and ket forms of the operator $\Oirr$ defined in 
Eq.(\ref{eq:NMnirrep}). However, before we start, we should make
sure that no problems occur with mixing, either for the defect
line considered in this paper, or for the parallel calculation
in the q-state Potts model with bulk disorder in the bond strength.

\subsection{Mixing With Other Operators}

\noindent The combinations $\Oirr$ of Eq.(\ref{eq:NMnirrep}) diagonalize the 
operators of the form $\prod\phi_{12}^\alpha$. However, we also
need to check that these operators don't mix with other operators 
that have the same scaling dimensions at $m=\infty$. 
We only need to consider operators of the form $\phi_{1q}$,
because these operators form a closed subalgebra in the minimal models.
To 2 loops, only 4 $\phi_{12}$ 
operators can be affected by the perturbation, 
and $4\;\mathrm{dim}(\phi_{12}) = 2$
at $m=\infty$. 
It easy to see that the only operator, or descendent of an operator,
or combination of operators, in the $\phi_{1q}$ subalgebra with 
dimension 2, is $\phi_{13}$ (therefore, to this order, no mixing occurs
with derivative operators).
$\phi_{13}$ does in fact mix with $(\phi_{12})^4$, and so
$\phi_{13}(\phi_{12})^{N-4}$ mixes with $(\phi_{12})^N$. However,
the 2-loop overlap integral is

\begin{eqnarray}
\nonumber
\lefteqn{ \sepri{2}{1} dx_2\sepri{3}{1} dx_3 
<\phi_{12}(x_1)\phi_{12}(x_2)\phi_{13}(\infty)>_0 } \\
 & & \hspace{1.08in} \twoptcorr{1}{2}\left(\twoptcorr{1}{3}\right)^2
\end{eqnarray}

\noindent This has no poles in $\eps$, so we would only need to 
take this mixing into account to do 3-loop calculations. 

Mixing is potentially more of a problem when we use these calculations
to get moments of the energy operator in the q-state Potts model with
bulk bond disorder -- this model is analyzed by expanding
about the Ising Model~\cite{dots:RG, Ludwig:irreps}, so we need to 
look for mixing with other operators that have the same scaling
dimension at $q=2$. Denoting the energy operator in the $\alpha$
replica by $\ve^{\alpha}$, we see that
$\ve^{\alpha}\ve^{\beta}\ve^{\gamma}$
and $\nabla^2\ve^{\alpha}$ have the same dimension
at $q=2$, and that in general we can replace any two $\ve$'s
with a $\nabla^2$ without changing the dimension. These operators
mix to two loops, with overlap integrals
such as

\pagebreak

\begin{eqnarray}
\nonumber
\int d^2x_2 \hspace{-0.22in} & & d^2x_3 \ \ve^\alpha \ve^\beta \ve^\gamma (0)
\ \ve^\alpha \ve^\beta (x_2) \ \ve^\alpha \ve^{\gamma} (x_3)
\ \nabla^2 \ve^{\alpha}(\infty) \\
\nonumber
& & = \int d^2x_2 d^2x_3 <\ve (0)\ve (x_2)\ve(x_3)\nabla^2\ve (\infty)>_0 \\
& & \hspace{0.9in} <\ve (0)\ve(x_2)>_0 <\ve (0)\ve (x_3)>_0 
\end{eqnarray}

This means at the combinations $\mathcal{O}_{NMn}$ 
of Eq.(\ref{eq:NMnirrep}) (with
$\phi_{12}$ replaced by $\ve$) are no longer generally
multiplicatively renormalizable, but mix with derivatives of
$\mathcal{O}_{N-2,\tilde{M},n}$. Luckily, however, no problem
with mixing occurs for the operators $\mathcal{O}_{N,N,n}$
and $\mathcal{O}_{N,N-1,n}$, which we know from the one-loop
calculations to provide the leading and subleading decay 
exponents. No problem with mixing occurs for $\mathcal{O}_{N,N,n}$
because the overlap

\begin{eqnarray}
\disopd{\alpha}{\beta}\disopd{\gamma}{\delta}|N,N,n>
\rightarrow \nabla^2 |N-2,M,n>
\end{eqnarray}

\noindent is 0 for all $M$, where the $\nabla^2$ is understood as
acting on only one of the $\ve$'s. It is not hard to see this
by counting the connected contractions of

\begin{eqnarray}
<N,N,n| \disopd{\alpha}{\beta}\disopd{\gamma}{\delta} |N-2,M,n>
\end{eqnarray}

\noindent If we expand $|N-2,M,n>$ into monomials, each monomial term
gives a contraction of 0 -- this is because after contracting
all the $\ve$'s in the monomial term, we will be left
with at least two antisymmetric terms from the $<N,N,n|$, and
thus positive and negative contractions will cancel. So the contractions
give a total of 0 (even before the replica limit $n\rightarrow 0$ is 
taken). This
is true even if the $|N-2,M,n>$ has a $\nabla^2$ on it, and even if
$<N,N,n|$ is replaced with $<N,N-1,n|$. So mixing is not a problem
for the $\mathcal{O}_{N,N,n}$ and $\mathcal{O}_{N,N-1,n}$
operators in the q-state Potts model with bulk bond disorder.

\subsection{First Order}

\noindent

\noindent Returning to our defect model, 
the $\first$ order correction comes from

\begin{equation}
\left(\disorderop\right)\ketirr\longrightarrow 
\OPEb\ketirr
\end{equation}

\noindent where we contract one $\phi_{12}$ operator in the 
$\disorderop$ with one $\phi_{12}$ operator in the
$\ketirr$. The integral is the same integral done in
Eq.(\ref{eq:A2.int}), giving $2\frac{r^{\eps}}{\eps}$. We
can get the combinatorial factor $\OPEb$ by contracting with
$\brairr$ on both sides :

\begin{equation}
\OPEb =\frac{\brairr\disorderop\ketirr}
   {\brairr \left. M,N,n\right\rangle}
\end{equation}

\noindent This combinatorial factor is found in subappendix~\ref{app-OPEb}.
Leaving our result in terms of $\OPEb$, 
the order $\Delta_0$ term in $Z_{NMn}$ is

\begin{equation}
2\OPEb\left(\frac{r^{\eps}}{\eps}\right)\Delta_0
\end{equation}

\subsection{Second Order}
\label{sec-secondRG-NMn}

\noindent To second order, we need to look at the number of 
ways that we can take

\begin{equation}
<N,M,n\mid_{\mu}\disopc{\alpha}{\beta}\disopc{\gamma}{\delta}
\rightarrow <N,M,n\mid_{\nu}
\end{equation}

\noindent The index $\mu$ in $<N,M,n\mid_{\mu}$ is a label used to make
discussion easier, and does not signify an independent variable.
The $\mu$ simply indicates that for discussing contractions,
the dummy indices $\alpha_i$ in Eq.(\ref{eq:NMnirrep}) will all be called
$\mu_i$ for $i=1,2\ldots n$. 

There are now a number of possible contractions
to consider. The one with
$\mu_i=\alpha=\gamma=\nu_j$ and $\mu_k=\beta=\delta=\nu_l$ (for
some $i$,$j$,$k$,$l$)
gives an integral over a squared four-point function.
We did this integral for $A_{33}$ in appendix~\ref{app-pertD}
and found that it had no singularities. So we can ignore this term.

We now consider the contraction with
$\mu_i=\alpha=\gamma=\nu_j$ and $\beta=\delta$, which
gives a combinatorial factor that we call $\Q{2}$, and the 
contraction with $\mu_i=\beta=\gamma=\nu_j$,
$\mu_k=\alpha$, and $\delta=\nu_l$, which gives a combinatorial
factor that we call $\Q{3}$. It is to be understood that no
replica indices other than the specified ones are equal to each other.
In both cases, we get the same integral as in
Eq.(\ref{eq:I2}) (or Eq.(\ref{eq:I2.alt})), which was found to be
$\mathcal{I}_2 = -2r^{2\eps}(\frac{1}{\eps}-\frac{2}{\eps^2})$.

To get the combinatorial factor for $\Q{2}$,
we look at

\begin{equation}
\frac{{}_{\mu}\hspace{-0.07in}
<N,M,n\mid\disopc{\alpha}{\beta}\disopc{\gamma}{\delta}
\mid N,M,n>_{\nu}}{<N,M,n\mid N,M,n>}
\end{equation}

\noindent and only allow the contractions described above.
For the contractions allowed in $\Q{2}$, the replica indices
that appear in a monomial term must be the same in the bra and
ket sides. The $\alpha=\gamma$ replica index can be
the same as any of the these $N$ monomial terms. The
$\beta=\delta$ replica index must be something different,
so has $(n-N)$ choices. We also get an overall factor of 4 for the
different pairs of terms that we could have chosen to contract.
So $\Q{2}=4N(n-N)$.

$\Q{3}$ can be calculated similarly. The 
$\alpha$ index contracts to the left and the $\delta$ index
contracts to the right  -- this gives a combinatorial factor
$<N,M,n\mid\sum_{\alpha\neq\delta}^{n} \phi_{12}^{\alpha}\phi_{12}^{\delta} 
\mid N,M,n>=\OPEb <N,M,n\mid N,M,n>$
The $\beta=\gamma$ index can then be any of the $(N-1)$ replica
indices in the right (or left) $\mid N,M,n>$ (excluding the
two which are equal to $\alpha$ or $\beta$). And there is
again an overall factor of 4, so
$\Q{3}=4(N-1)\OPEb$. 

Now consider the contraction with
$\mu_i=\alpha$, $\beta=\gamma$ and $\delta=\nu_j$,
which gives a combinatorial factor $\Q{4}$, and the
contraction with $\mu_i=\alpha$, $\mu_j=\gamma$, 
$\nu_i=\beta$, and $\nu_j=\delta$, which gives a 
combinatorial factor $\Q{5}$. Both of these terms
give the same integral as in Eq.(\ref{eq:I1}) (or Eq.(\ref{eq:I1.alt})),
and thus a factor of 
$\mathcal{I}_1=4\frac{r^{2\eps}}{\eps^2}$ .

$\Q{4}$ can be evaluated in the same manner as
$\Q{3}$, giving $\Q{4}=4(n-N-1)\OPEb$.
The term $\Q{5}$ is more complicated and is
found in subsection~\ref{sec-Q5.c} to be
$\Q{5}=-4N(n-N)-2(n-2)\OPEb+(\OPEb)^2$. Putting this all 
together, we get the result

\begin{eqnarray}
\nonumber
Z_{NMn} = 1+2\OPEb\frac{r^{\eps}}{\eps}\Delta_0+
	\frac{1}{2}(\Q{2}+\Q{3})
	(-2\frac{r^{2\eps}}{\eps}+4\frac{r^{2\eps}}{\eps^2})+ \\
\nonumber
	\frac{1}{2}(\Q{4}+\Q{5})
	(4\frac{r^{2\eps}}{\eps^2})+\mathcal{O}(\Delta_0^3) \\
\nonumber
 = 1 + 2\OPEb\frac{r^{\epsilon}}{\epsilon}\Delta_0
   -4\left(N(n-N)+(N-1)\OPEb\right)\frac{r^{2\epsilon}}{\epsilon}\Delta_0^2 \\
   + \left(2(\OPEb)^2+4(n-2)\OPEb\right)
     \frac{r^{2\epsilon}}{\epsilon^2}\Delta_0^2
   +\mathcal{O}(\Delta_0^3)
\end{eqnarray}


\newpage
\section{Combinatorial Factors}
\label{app-combo}

\noindent In this appendix we calculate the combinatorial
factors used in appendix~\ref{app-RGNM}. We want the number of 
possible contractions between the irreducible representations
of $S_N$, given by $\Oirr$ in Eq.(\ref{eq:NMnirrep}),
and copies of the disorder operator. The trivial
spatial dependence is supressed in the equations below. 


\subsection{Normalization}

\noindent First, we calculate the normalization of $\mid N,M,n>$

\begin{eqnarray}
\nonumber
A[N,M,n] & \equiv &  <N,M,n \mid N,M,n>=  
\irrsum{\alpha}{M}\irrsum{\beta}{M}  \\
\nonumber
 & & \hspace{-0.155in} <\irrterm{\alpha}{M}{N}\mid\\
 & & \irrterm{\beta}{M}{N}>
\end{eqnarray}

\noindent If we expand out the two terms
$(\phi_{12}^{\alpha_M}-\phi_{12}^{n-(M-1)})$ and
$(\phi_{12}^{\beta_M}-\phi_{12}^{n-(M-1)})$ into
monomials, the cross terms are 0, and the remaining terms are
of the same form $A[\ldots]$ as before, but with different values of
$M$, $N$ and $n$. We get

\begin{equation}
A[N,M,n]=A[N,M-1,n-1]+(n-M-N+1)^2 A[N-1,M-1,n-1]
\end{equation}

\noindent We can calculate $A[N,M,n]$ explicitly when $M=0$ :

\begin{eqnarray}
\nonumber
A[N,0,n] & = & \irrsumb{\alpha}\irrsumb{\beta}
   <\phi_{12}^{\alpha_1}\phi_{12}^{\alpha_2}\ldots
    \phi_{12}^{\alpha_N} \mid \phi_{12}^{\beta_1}
    \phi_{12}^{\beta_2}\ldots\phi_{12}^{\beta_N}> \\
 & = & N!\prod_{i=1}^{N} (n-N+i)
\end{eqnarray}

\noindent Here the $N!$ comes from the number of ways to contract
the left side with the right side, and the factors of $n-N+i$ come from
the different ways to pick the $\alpha_i$ once the contractions
have been done. Given the value of $A[N,M,n]$ when $M=0$, and the
recursion relation above, we can show by induction that the
general solution is

\begin{equation}
A[N,M,n]=(N-M)![\prod_{i=1}^N (n-M-N+i)]
   [\prod_{i=1}^M (n-2M+1+i)]
\end{equation}


\subsection{OPE -- 1st order term}
\label{app-OPEb}

\noindent Here we want to calculate 

\begin{equation}
B[N,M,n]\equiv <N,M,n\mid\disopc{\alpha}{\beta}\mid N,M,n>
\end{equation}

\noindent Note that three-point functions of $\phi_{12}$ are
$0$ by the $\phi_{12}\rightarrow -\phi_{12}$ duality 
\linebreak symmetry. 
As before, we expand out the 
$(\phi_{12}^{\alpha_M}-\phi_{12}^{n-(M-1)})$ and
$(\phi_{12}^{\beta_M}-\phi_{12}^{n-(M-1)})$ terms into
monomials. The direct terms again give contractions of the
form $B[\ldots]$, but we now have a cross term in which
either $\alpha$ or $\beta$ must be equal to
$n-(M-1)$. We get

\begin{eqnarray}
\nonumber\frac{1}{2} B[N,M,n] & = & \frac{1}{2}B[N,M-1,n-1] \\
\nonumber
 & & \hspace{-0.7in} +(n-M-N+1)^2 \frac{1}{2} B[N-1,M-1,n-1]  \\
\nonumber
 & & \hspace{-0.7in} - 2(n-M-N+1)\irrsum{\alpha}{M}\irrsum{\beta}{M}\sum_{\alpha=1}^{n}\\
\nonumber
 & &  \hspace{-0.44in} <\phtwo{\alpha_1}{n}...\phtwo{\alpha_{M-1}}{n-(M-2)}
	\ph{\alpha_M}\ph{\alpha_{M+1}}...\ph{\alpha_N} 
   \mid\ph{\alpha}\mid \\
 & & \hspace{-0.36in} \mid\phtwo{\beta_1}{n}...\phtwo{\beta_{M-1}}{n-(M-2)}
  \hspace{.28in}\ph{\beta_{M+1}}...\ph{\beta_N}>
\end{eqnarray}

\noindent The contraction in the last term looks just like the contraction
$A[N,M-1,n-1]$, with $\alpha$ taking the place of $\beta_M$. The only 
difference between the sum above and $A[N,M-1,n-1]$, is that 
the sum over $\alpha$ above includes terms with $(n-M+2)\leq\alpha\leq n$, 
and terms where $\alpha$ is the same as some other term $\beta_i$, for 
$1\leq i\leq (M-1)$. However, these two extra contributions are
equal and opposite, being equal to $\pm(M-1)(n-M-N+1)A[N-1,M-2,n-2]$. So
they cancel, and we have

\begin{eqnarray}
\nonumber
\frac{1}{2}B[N,M,n] & = & \frac{1}{2}B[N,M-1,n-1] +  \\
\nonumber
 & & (n-M-N+1)^2 \frac{1}{2} B[N-1,M-1,n-1] -\\
 & & 2(n-M-N+1)A[N,M-1,n-1]
\end{eqnarray}

\noindent As with $A[N,0,n]$, it is easy to calculate $B[N,0,n]$ :

\begin{equation}
B[N,0,n]=2N(N!)[\prod_{i=0}^N (n-N+i)]=2N(n-N)A[N,0,n]
\end{equation}

\noindent Given the recursion relation for B[N,M,n], the initial
condition B[N,0,n], and the result for A[N,M,n] in the previous
section, we can show by induction that

\begin{equation}
\label{eq:bbar}
\OPEb\equiv\frac{B[N,M,n]}{A[N,M,n]}
 = 2((N-M)n-N^2+M(M-1))
\end{equation}


\subsection{2nd order term}
\label{sec-Q5.c}

\noindent We want to calculate

\begin{equation}
C[M,N,n] \equiv 4 <N,M,n\mid\disopc{\alpha}{\beta}
   \disopc{\gamma}{\delta}\mid N,M,n>
\end{equation}

\noindent where we require that $\alpha$ and $\gamma$ contract
to the left, the $\beta$ and $\delta$ contract to the right,
and $\alpha$, $\beta$, $\gamma$, $\delta$ are all distinct from
one another (other possible directions of 
contractions give the factor of 4 in front).
This combinatorial factor arises in the contraction 
for $\Q{5}$ in subappendix~\ref{sec-secondRG-NMn}
As in the previous subsection, expanding out the
$(\phi_{12}^{\alpha_M}-\phi_{12}^{n-(M-1)})$ and
$(\phi_{12}^{\beta_M}-\phi_{12}^{n-(M-1)})$ terms into
monomials gives back two terms of the form $C[\ldots]$, and
a more complicated cross-term. The cross term almost has the same 
form as $B[N,M-1,n-1]$, 
but we also get some extra terms because 
the sums over $\alpha$, $\beta$, $\gamma$ and $\delta$ are
different than we would have in a $B[\ldots]$ term.  Counting all the
ways in which our cross term differs from $B[N,M-1,n-1]$ is
tedious but straightforward, and we get

\begin{eqnarray}
\nonumber
\lefteqn{\hspace{-0.2in}
	 \frac{1}{2}B[N,M-1,n-1]-(M-1)(n-M-N+1)^2A[N-1,M-2,n-2] } \\
\nonumber
 & & -(M-1)A[N,M-2,n-2] -(n-M-N)A[N,M-1,n-1] \\
 & & = \frac{1}{2}B[N,M-1,n-1]-(n-N-1)A[N,M-1,n-1]
\end{eqnarray}

\noindent The recursion relation is

\begin{eqnarray}
\nonumber
\lefteqn{ \frac{1}{4} C[N,M,n] = \frac{1}{4}C[N,M-1,n-1] + }\\
\nonumber
 & & (n-M-N+1)^2\frac{1}{4}C[N-1,M-1,n-1] - 4(n-M-N+1) \\
 & & \left\{\frac{1}{2}B[N,M-1,n-1]-(n-N-1)A[N,M-1,n-1]\right\}
\end{eqnarray}

\noindent Combined with the initial condition,

\begin{equation}
C[N,0,n]=4N(N-1)(n-N)(n-N-1)A[N,0,n],
\end{equation}

\noindent we can find the value of $C[N,M,n]$ for all $M$ and $N$ by
induction. The result is

\begin{equation}
\Q{5}\equiv\frac{C[N,M,n]}{A[N,M,n]} 
 = -4N(n-N)-2(n-2)\OPEb+\left(\OPEb\right)^2
\end{equation}

\noindent where $\OPEb$ was defined in Eq.(\ref{eq:bbar}).


\newpage
\section{Ising Ladder Defect for \protect\boldmath$K<0$}
\label{app-IsingLadder}

The branch of the arctangent used in Eq.(\ref{eq:glambdaladder})
for antiferromagnetic ($K=K_c+\lambda <0$) ladder couplings
requires some explanation. We take the value of the arctangent to be 
in $(0,\frac{\pi}{2})$ if its argument is positive (i.e. $K>0$) and to be
in $(\frac{\pi}{2},\pi)$ if its argument is negative (i.e. $K<0$).
This makes the slope of $g(\lambda)$ continuous through $K=0$. On the
other hand, the results of~\cite{Bariev,McCoy} have a slope discontinuity
at $K=0$, and have g symmetric under $K\rightarrow -K$. This 
slope discontinuity results from a 
level crossing in the lowest scaling dimension for operators on the
boundary~\cite{Affleck}. If we let $\sigma_t$ be the spin on one side of
the defect, and $\sigma_b$ the spin on the other side, we see that the 
operator with the lowest scaling dimension changes from $\sigma_t + \sigma_b$
to $\sigma_t - \sigma_b$ as $K$ goes through $0$, so that while the 
dimension of each operator changes smoothly through $K=0$, the dimension of
the lowest scaling operator does not. However, if we take our random applied 
magnetic field to not vary across the defect, it couples to 
$\sigma_t + \sigma_b$ only, and we want the lowest scaling dimension 
for $K>0$, but the $\second$ lowest scaling dimension for $K<0$.
We can get these from~\cite{Affleck}, thus justifying the branches of 
arctangent chosen above. 

Note that if we had used the other branch of the arctangent, 
corresponding to a magnetic field uncorrelated across the defect, the 
flow picture in the ladder case would have been symmetric under $K\rightarrow -K$, 
and the entire $(\Delta,\lambda)$ plane would have flowed into the decoupled 
point. 


\pagebreak

\pagebreak


\newpage

\noindent {\Large\bf Figure Captions}

\bigskip

\noindent Fig. 1. \ \ A two-point correlation function for operators
lying along the defect.

\bigskip

\noindent Fig. 2. \ \  A one-point function for an operator in the
bulk.

\bigskip

\noindent Fig. 3. \ \ The ladder defect in the Ising model.

\bigskip

\noindent Fig. 4. \ \ The chain defect in the Ising model.

\bigskip

\noindent Fig. 5. \ \ Renormalization Group flows for the ladder
defect in the Ising model. $\Delta$ is the strength of the disordered
magnetic field along the defect line, and $\lambda$ is the bond
strength along the line.

\bigskip

\noindent Fig. 6. \ \ Renormalization Group flows for the chain
defect in the Ising model.


\newpage

\begin{picture}(120,60)
\put(44,50){Figure 1}
\put(0,20){\line(1,0){120}}
\put(7,35){$(x_1,0)$}
\put(87,35){$(x_2,0)$}
\put(20,25){\circle*{3}}
\put(100,25){\circle*{3}}
\end{picture}

\bigskip
\bigskip
\bigskip
\bigskip

\begin{picture}(120,120)
\put(40,100){Figure 2}
\put(0,10){\line(1,0){120}}
\put(60,80){\circle*{3}}
\put(65,83){$(0,y)$}
\multiput(60,10)(0,10){7}{\line(0,1){5}}
\put(80,45){$y$}
\put(70,20){\line(0,1){20}}
\put(70,50){\line(0,1){20}}
\qbezier(60,80)(67,80)(70,70)
\qbezier(60,10)(67,10)(70,20)
\qbezier(70,40)(70,45)(77,45)
\qbezier(70,50)(70,45)(77,45)
\end{picture}

\bigskip
\bigskip
\bigskip

\setlength{\unitlength}{0.9pt}
\begin{picture}(350,200)
\put(38,160){Figure 3}
\put(243.5,160){Figure 4}
\thinlines
\multiput(0,40)(0,10){11}{\line(1,0){110}}
\multiput(10,30)(10,0){10}{\line(0,1){120}}
\multiput(200,40)(0,10){11}{\line(1,0){120}}
\multiput(210,30)(10,0){11}{\line(0,1){120}}
\linethickness{1.7pt}
\multiput(50,40)(0,10){11}{\line(1,0){10}}
\put(260,30){\line(0,1){120}}
\end{picture}
\setlength{\unitlength}{1.0pt}

\pagebreak

\begin{figure}[tb]
\begin{center}
\epsfbox{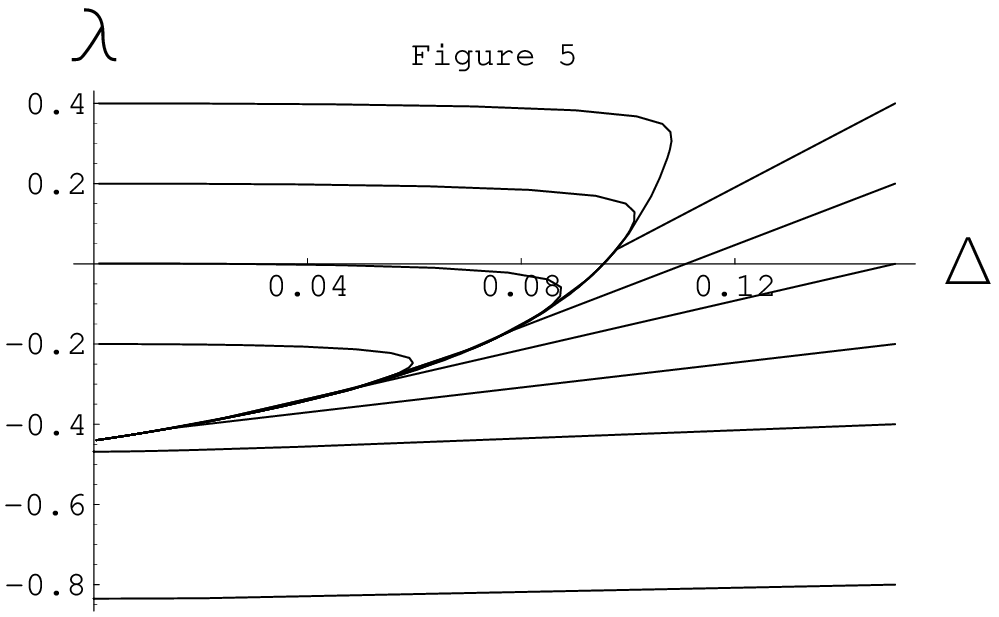}
\end{center}
\end{figure}

\begin{figure}[tb]
\begin{center}
\epsfbox{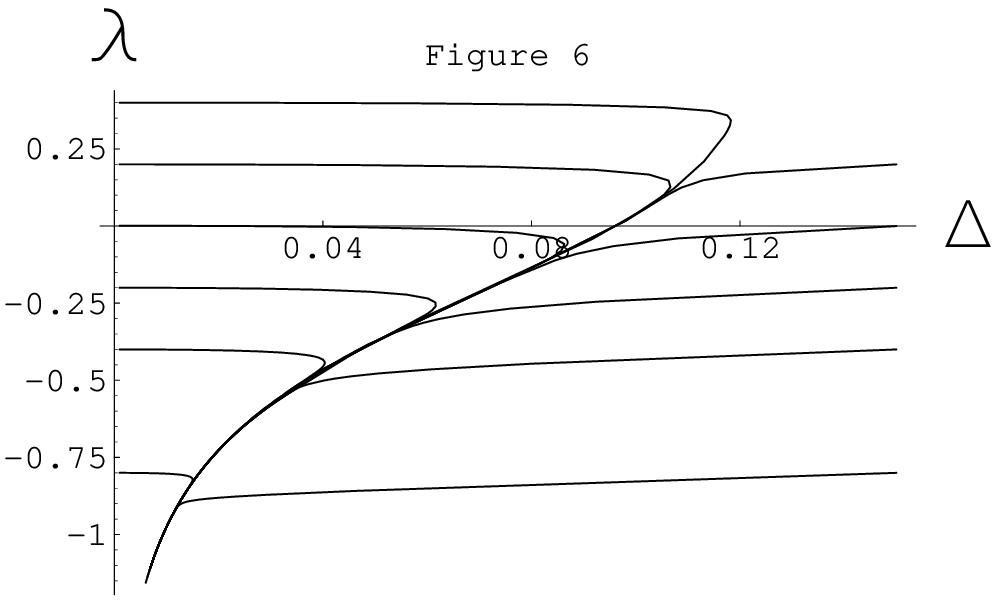}
\end{center}
\end{figure}

\end{document}